\documentclass{vgtc}                          

\graphicspath{{figures/}{pictures/}{images/}{./}} 

\usepackage{times}                     

\usepackage{tabu}                      
\usepackage{booktabs}                  
\usepackage{lipsum}                    
\usepackage{mwe}                       

\usepackage{mathptmx}                  

\usepackage{graphicx}
\usepackage{amsmath}
\usepackage{amssymb}
\usepackage{paralist}
\usepackage{tabularx}
\newcolumntype{C}{>{\arraybackslash}X}

\onlineid{0}

\vgtccategory{Research}

\vgtcinsertpkg

\title{Bringing physics-based controller policies to virtual reality}

\author{Joan Llobera\thanks{e-mail: joan.llobera@artanim.ch}\\ %
        \scriptsize Artanim Foundation%
\and Pierre Nagorny\\ %
     \scriptsize Artanim Foundation%
\and Balint Hodossy\\ %
     \scriptsize Imperial College London%
\and Valérie Juillard\\ %
     \scriptsize Artanim Foundation%
\and Caecilia Charbonnier\\ %
     \scriptsize Artanim Foundation%
     }

\teaser{
  \centering
  \includegraphics[width=0.6\textwidth]{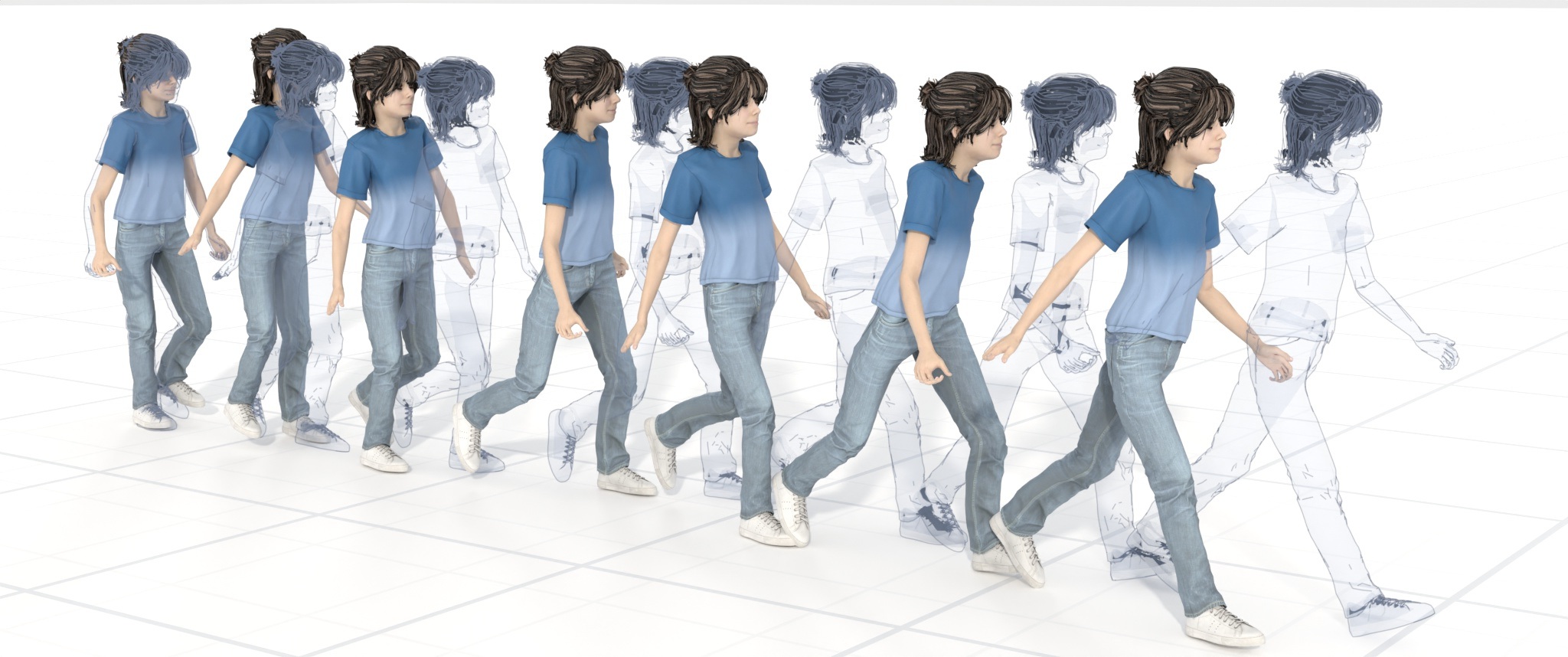}
  \caption{Character driven by physics-based controller (solid), compared to the reference animation (transparent). 
  Snapshots sampled at
  0.5\,s intervals. 
  }
  \label{fig:physics_based_controller}
}

\abstract{

The use of physics-based controllers in video games and virtual reality (VR) is now technically possible, but little attention has been paid to evaluating the movement quality generated by such controllers, particularly when high-quality motions are needed. Moreover, these controllers have been developed in research-oriented physics engines, which are not suitable for production projects, where game engines like Unity3D or Unreal Engine are preferred.
Nowadays, it is easy to set up any element of a VR experience so users can interact with it physically, except for interactive VR characters. Can we extend this to animated VR characters?

In this work, we studied whether physics-based character animation (PBCA) controllers can produce high-quality movement in VR-compatible environments (Unity3D, WebXR).  
Since VR experiences typically use rigged and textured characters, we evaluated the quality of the movement by capturing and manually retargeting eight walking styles 
to two rigged and textured characters, which served as reference animations for comparison with the motions generated by the PBCA controllers.
We also found that showing the training results on rigged and textured characters helped highlight the controller policies' animation inaccuracies.

We evaluated the quality of the results, as well as the losses at each stage of the processing pipeline: data ingest, training and inference.  For data ingest, we computed accuracy metrics related to the retargeting method used.
For training, we computed accuracy metrics related to different training policies and different training environments. 
We developed new JAX implementations of state-of-the-art PBCA controllers (\emph{DeepMimic}, \emph{AMP} and \emph{ADD}) which can be trained in MuJoCo-MJX in minutes, 
50 times faster than in a traditional CPU-based training environment. The benefit of these particular implementations is that the resulting control policies 
can be readily transferred to Unity3D and support cross-platform compilation, including Android, something critical for use in VR production. 
For inference, we tested these controller policies in Unity3D and WebXR. 
Our results suggest that \emph{DeepMimic} and \emph{ADD} trained in MuJoCo-MJX generate controller policies with better animation fidelity than other methods. 

To simplify the comparison between controller policies, and different reference animations, we implemented an animation fidelity index (\emph{afi}). 
Its results are analogous to other more common metrics, but \emph{afi} is more intuitive to interpret and thus simplifies the comparison of training methods across the successive stages of the processing pipeline. 
We also did an expert comparison between the reference motions and the end result.

Our results suggest that any use of PBCA in production projects should be approached with caution. Some inaccuracies readily visible to an animator were difficult to detect with the different metrics commonly used, 
and the overall quality of the results varied considerably depending on different choices made across the processing and delivery pipeline. 
Finally, we provide the animation data and code to facilitate the transfer of physics-based control research results to the artistic and production communities.

} 


\begin{document}


\firstsection{Introduction}

\maketitle

Since DeepMimic~\cite{peng_deepmimic:_2018} and DReCon~\cite{bergamin_drecon_2019} demonstrated that physics-based controllers driven by Reinforcement Learning (RL) policies can learn to mimic reference motion capture clips, 
a critical question remained underexplored:  can we use physics-based controller policies to drive virtual reality (VR) characters? The answer seems to be \emph{sometimes, but unfortunately not if we want to closely match high-quality reference animations}. 

The lack of a standard integration of research-oriented physics engines into production-oriented game engines remains a barrier. 
It is also necessary to check the movement quality generated by the controller policies at each stage of the pipeline to transfer the controllers in a production environment. 
This can partially be done with objective metrics, but it is also necessary to rely on the expertise of an animation specialist. 

In character animation, the movement \emph{is} the means of expression, and any degradation in fidelity directly undermines the artistic intent. 
However, most evaluations focus on task success---survival rate, tracking reward, perturbation recovery--- and rarely measure the fine-grained animation quality that determines production readiness. 
Moreover, results are typically demonstrated in research simulators rather than in production game engines (e.g., Unity3D, Unreal Engine) where animation content is actually deployed.

In this work, we implemented an end-to-end pipeline to train controller policies from an animated character, and a systematic evaluation of animation fidelity at different stages of the pipeline. 
Following the trends in the Physics-based character animation (PBCA) community, our pipeline trained on GPU, which is much faster than CPU training. 
However, to meet the requirements of production deployments, the inference was done on the CPU in VR-compatible environments (Unity3D, WebXR). 
We evaluated the quality of the outcome of three physics-based controller policies (\emph{DeepMimic} \cite{Peng:2018:DeepMimic}, \emph{AMP} \cite{Peng:2021:AMP}, \emph{ADD} \cite{zhang_add_2025}), 
and compared it with the outcome of a simpler and slower CPU-based training pipeline.

To assess the flexibility and reliability of different controller policies, we tested each of them on two characters and eight walking styles. 
We chose walking reference animations because they should be relatively easy for a PBCA policy to learn. 
This allowed us to focus our evaluation on animation quality, something critical for the adoption of these techniques in a production context.
In summary, our contributions are:
\begin{enumerate}
\item A GPU-accelerated pipeline with JAX implementations of state-of-the-art PBCA controller policies (\emph{DeepMimic}, \emph{AMP}, and \emph{ADD}). From an  FBX animation file, this pipeline
trained a policy in minutes on a single GPU, converted it to ONNX and integrated it into a Unity3D scene, ready for cross-platform deployment, including Android-based VR headsets. 
We also deployed the controller policies on a VR-compatible web environment based on WebXR. 
\item A high-quality animation benchmark comprising two rigged and textured characters of different body proportions (adult male and female child), each performing eight walking styles, with reference motions manually retargeted to the characters by a professional animator.
\item An evaluation of animation fidelity losses for different policies, each trained on different reference animations. We evaluated animation fidelity
across all stages, from the source animation through training and to the final deployed controller policy. 
This was done with commonly used metrics, as well as with a new animation fidelity index, and visual inspection of each animation-policy pair. 
We also compared the policies with the outcome of an earlier CPU-based training pipeline that is also VR-compatible. 

\end{enumerate}

\section{Background}

\subsection{Physics engines for VR production}

The two most popular game engines on the market are currently Unity3D\footnote{https://unity.com/} and Unreal Engine\footnote{https://www.unrealengine.com}, which have considerably democratized video game and VR production.
These game engines provide integrated physics engines (PhysX and Chaos for Unity3D and UE5, respectively) to simulate behaviours of physical entities. While their capabilities are suitable for many use cases, they were not designed to meet the needs of solving optimal control problems. They provide limited access to the physics engine API, and they are not compatible with modern training pipelines that are based on GPU-accelerated physical simulations integrated in the deep learning training stack.
Moreover, research-oriented engines (MuJoCo\footnote{https://mujoco.org/}, Bullet\footnote{https://github.com/bulletphysics/bullet3}, DART\footnote{https://dartsim.github.io/}) and production game engines differ in their collision handling and numerical precision, and the extent to which policies trained in one physics engine can transfer to another remains unclear. 
We therefore choose to take advantage of the MuJoCo plugin for Unity3D, which allows evaluating and comparing our two training pipelines inside the same development environment.

WebXR allows for the rendering of interactive VR environments in web browsers and is emerging as an alternative to traditional game engines for VR development. We also evaluate how it holds up against a standalone VR build.

\subsection{Humanoid controllers}

Humanoid controllers broadly divide into \emph{tracking-based controllers}, which combine a kinematic motion planner with a physics-based controller trained to imitate the poses generated by the kinematic motion planner,  and \emph{sensorimotor controllers}, where the policy directly maps perceptions to motor actions through reward-driven learning~\cite{Llobera:2023:PBCA}.  

\subsubsection{Tracking-based controllers}

Tracking-based controllers can produce high-quality animation by blending pre-recorded clips at runtime. However, the interaction with the environment is driven by the kinematic poses that the physical controller is trained to imitate, and they therefore lack physical interaction with the environment.
DReCon~\cite{bergamin_drecon_2019} combines a physics-based controller with a kinematic motion planner~\cite{buttner_motion_2015,holden_learned_2020}. At inference time, the trained policy receives both the current physics state and a kinematic reference pose, and outputs joint torques that track the sequence of kinematic poses while maintaining balance.
\ \\
While simplifying the training process, this method involves additional computational overhead at inference time due to the need for kinematic motion synthesis and to provide an expanded observation space.
\subsubsection{Sensorimotor controllers with explicit rewards}

DeepMimic \cite{peng_deepmimic:_2018} demonstrated that carefully designed tracking rewards---exponentially weighted errors on joint positions, velocities, center of mass, and end-effectors---enable RL policies to imitate reference motions.
This approach produces high fidelity on single clips but requires per-component reward tuning and explicit tracking of the animation clip's phase (from 0 to 1) at inference time.
VMP~\cite{Serifi:2024:VMP} extends explicit rewards to multi-motion settings by decoupling motion understanding (a pre-trained Variational Autoencoder on short motion windows) from physics control (a conditional RL policy), achieving stable convergence on 11 hours of diverse motion data.
AMOR~\cite{Alegre:2025:AMOR} further extends this by training a single multi-objective policy conditioned on reward weight vectors, enabling post-training trade-off tuning.
The Disney Research robotics pipeline~\cite{Grandia:2024:BDX} similarly relies on DeepMimic-style rewards for production deployment of bipedal character robots, confirming that explicit rewards remain the preferred approach when animation fidelity is the priority.

\subsubsection{Sensorimotor controllers with adversarial rewards}

AMP~\cite{peng_amp_2021} replaces hand-crafted tracking rewards with a learned discriminator that distinguishes reference motion transitions from policy-generated ones, using a GAIL-style adversarial objective.
This eliminates reward engineering and captures stylistic features that are hard to specify analytically.
Subsequent extensions include ASE~\cite{Peng:2022:ASE} (reusable skill embeddings), CALM~\cite{Tessler:2023:CALM} (conditional latent models), and ReGAIL~\cite{Boursin:2024:ReGAIL} (relative features for single-motion GAIL).
ADD~\cite{zhang_add_2025} modifies the discriminator to operate on displacement vectors between reference and agent states, using the zero vector as the only positive sample.
MimicKit~\cite{Peng:2025:MimicKit} provides a unified framework implementing \emph{DeepMimic}, \emph{AMP}, and \emph{ADD} for fair comparison.
A recent survey~\cite{Li:2025:FeatureVsGAN} systematically compares feature-based and GAN-based approaches, concluding that explicit tracking rewards excel at high-fidelity single-motion imitation while adversarial methods offer better scalability but suffer from training instability and coarser reward signals.
\subsubsection{Scaling and universal controllers}
PHC~\cite{Luo:2023:PHC} uses progressive training and residual forces for robust multi-motion control.
PULSE~\cite{Luo:2024:PULSE} learns a universal 32-dimensional motion latent space from the AMASS dataset.
SuperPADL~\cite{Juravsky:2024:SuperPADL} scales to 5000+ skills via progressive distillation.
SONIC \cite{Luo:2025:SONIC} demonstrates that scaling data (100M+ frames), compute, and model size all improve motion tracking quality.
BeyondMimic~\cite{Liao:2025:BeyondMimic} achieves state-of-the-art tracking with unified hyperparameters across highly dynamic motions.
CLoSD \cite{Tevet:2025:CLoSD} combines a diffusion-based motion planner with a physics-based tracker in a closed feedback loop.
Recent contributions have scaled these approaches to thousands of motions~\cite{Juravsky:2024:SuperPADL, Luo:2025:SONIC,wang2026motionbricksscalablerealtimemotions}, enabled universal controllers~\cite{Luo:2024:PULSE, Liao:2025:BeyondMimic}, and transferred policies to physical robots~\cite{Grandia:2024:BDX, He:2025:ASAP}.

\subsection{Animation fidelity metrics}

\label{metrics}

Despite rapid progress in controller capabilities, quantitative evaluation of animation quality remains inconsistent across the literature.
Most papers report task-level metrics (survival rate, tracking reward, success percentage) rather than fine-grained fidelity measures.
Common metrics include joint position mean average error (MAE) in degrees~\cite{Serifi:2024:VMP}, end-effector position error in meters~\cite{Peng:2018:DeepMimic}, or motion success rate~\cite{Luo:2024:PULSE}.

The recent tracking literature (VMP~\cite{Serifi:2024:VMP},
CALM~\cite{Tessler:2023:CALM}, BDX~\cite{Grandia:2024:BDX} and
ASAP~\cite{He:2025:ASAP}) generally show a small set of
metrics that allow comparing the methods: joint Mean Average Error (\emph{MAE}) (in degrees or radians), root-relative Mean Per Joint Position Error (\emph{MPJPE})
(mm), end-effector \emph{MAE} (m), $e(n)$ (m/frame), and success rate
(\% of episodes that survive a fixed rollout horizon).

Perceptual studies have partially identified what the important factors are to judge pose similarity.
The set of quaternions defining a pose for a given skeleton is not considered a good measure of perceived pose similarity~\cite{tang2008emulating}. Indeed, even minor deviations early in the kinematic chain will lead to obvious deviations in the end effectors. Furthermore, the angular distance between the pose joints is unable to capture deviations in higher-order kinematics that affect both visual fidelity and stability.
The use of angular velocity for motion prediction seems more appropriate than the combination of pose and linear velocity \cite{harvey2020inbetweening}.
It is also easier to interpret than other proposed metrics, such as Neural Pose Spectrum Similarity (NPSS) \cite{gopalakrishnan2019neural}. The physical meaning is more intuitive, and it can more easily be visualised in a graph to compare a reference animation and movement produced by a controller.
For end-effectors, Euclidean distance has also been shown to matter, particularly for hands and fingers~\cite{harada2004, basset2022}.

\section{Materials and Methods}

\subsection{Reference Data}

In most of the literature on physics-based controllers, motion retargeting is performed automatically.
This is practical, but difficult to reconcile with high animation quality.
Our experience, further confirmed by the results presented below, suggests that manual retargeting performed by a skilled animator, combined with a ragdoll matching the proportions of the reference skeleton, is what introduces a smaller fidelity loss. 
This is consistent with the production pipeline in \cite{Grandia:2024:BDX}, where artist-created reference motions are used to ensure animation quality in deployed robotic characters.
Investing in manual retargeting, rather than relying on automated solutions, appears to be a sensible choice.

To build a controlled, high-quality dataset and allow testing the extent to which the training pipeline works for characters of different proportions,
we recruited two volunteers ---a 10-year-old girl (LL) and a 35-year-old man (MM)--- and instructed them to walk straight ahead in 8 different styles such as ``neutral'', ``angry'' and ``wounded'' (see list in Table \ref{table1:fid_scores}).
The motion was recorded using a Vicon MX T-Series motion capture system (Vicon, Oxford Metrics, UK) consisting of 24 cameras (24 × T40S) sampling at 120 Hz.
Vicon Shogun version 1.18 was used to generate a skeleton matching the actors’ proportions to automatically transfer each captured take to the target skeleton.

We then transferred the animation data to rigged and textured characters, in three steps.
First, a technical artist created a textured character model in
Character Creator version 4 (one character per actor).
Second, the Shogun skeleton was adapted using Blender to be
compatible with Unity3D's Humanoid definition--- the number of bones in the spine was
reduced from four to three to match the Humanoid layout, while preserving the actors' segment
lengths found in the Shogun reference, and each character skin was rigged onto this adapted skeleton.

Third, a professional 3D animator used Motion Builder version 2026 to
retarget each captured take from the Shogun skeleton onto the
Unity3D-compatible skeleton and to clean the result.
Since the motion-capture rig used a 3-finger marker setup that does not resolve every finger joint, the animator manually set each hand to a natural rest pose; no other body-part data was edited.
Each take was then trimmed to an animation cycle of four footsteps. The final animation files used to train the physics-based controllers are in the supplementary materials.

\subsection{Training}

We used two different training pipelines, \emph{playground} and \emph{modular-agents}.

\begin{figure}
  \centering
 \includegraphics[width=0.5\textwidth]{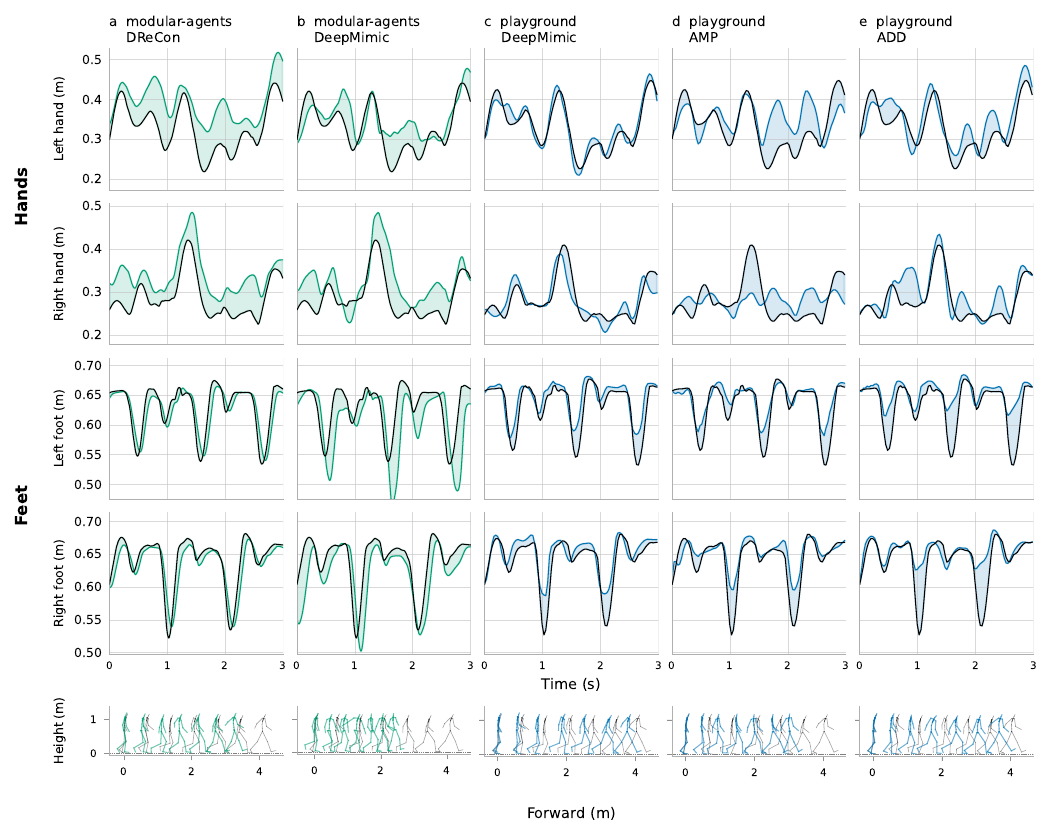}
\caption{End-effector distances to root for the Happy Walk (LL skeleton) and time-aligned corresponding sagittal stick figures. This visualisation allowed us to rapidly detect whether a method introduces foot skating or whether it reproduces faithfully an animation involving touching or grabbing an object.
}
\label{ee}
\end{figure}

\subsubsection{Playground}
\label{training2}

The main training pipeline used MJX~\cite{Zakka:2025:MuJoCoPlayground}, a JAX-compiled version of MuJoCo that runs entirely on GPUs.
The RL policy used was PPO, implemented using the Brax training library~\cite{Freeman:2021:Brax}, with 4096 parallel environments per training run.
The control policies trained were JAX reimplementations of \emph{DeepMimic}, \emph{AMP}, and \emph{ADD}.
Policies were exported as ONNX models and loaded in Unity3D via the ONNX Runtime plugin.
The main motivation for this training pipeline was to use modern hardware acceleration, 
as adopted in many recent papers on physics-based controllers~\cite{Zakka:2025:MuJoCoPlayground, Freeman:2021:Brax, Serifi:2024:VMP, Luo:2025:SONIC},
which yielded at least an order of magnitude in training speed.
This also enabled the adoption of the kind of massively parallelized training strategies recently demonstrated on motion-imitation benchmarks \cite{Luo:2023:PHC,tessler2024maskedmimic,Luo:2025:SONIC}.

This strategy introduced an additional challenge: the MJX (GPU) and MuJoCo~C implementations differ in their collision handling and numerical precision, introducing a sim-to-sim transfer gap analogous to the sim-to-real gap in robotics~\cite{He:2025:ASAP, Serifi:2025:Thesis}.
 We minimized this gap through two strategies: (a)~careful observation design---we excluded simulator-specific quantities such as composite inertia tensors (\texttt{cinert}) from the policy observations, as these differ between MJX and MuJoCo~C and neural networks are sensitive even to small discrepancies, and (b)~three-way validation comparing the policy output in the MJX training environment, in MuJoCo~C and in Unity3D, flagging any experiment where cross-platform metrics diverged beyond $2\sigma$.

To systematically compare different characters and animations, we developed an automated pipeline that created the entire training dataset---MJCF ragdoll, reference joint trajectories, and animation metadata---directly from the FBX animation file using inverse kinematics retargeting.\footnote{https://github.com/kevinzakka/mink}

 \subsubsection{Modular agents}
 \label{training1}
For comparison, we used the \emph{modular-agents} training pipeline. This pipeline  is based on extensions of ML-Agents in Unity3D. For training it uses MuJoCo compiled for CPU (MuJoCo~C) 
and integrated in a plugin for Unity3D.
 The RL policy used was Proximal Policy Optimization (PPO) \cite{Schulman:2017:PPO}, in particular the implementation available in the ML-Agents package (version 4.0.0)\footnote{https://github.com/unity-technologies/ml-agents}.
The training scene was implemented in Unity3D version 6.3.3f1, using a custom library developed to generate the training events and the RL signals (observations, rewards, actuators), which is publicly available~\cite{Hodossy:2024:ModularAgents}.
The training environment was generated by compiling a Linux binary of the Unity3D scene, communicating with the ML-Agents Python training package over a client-server socket.
The control policies used for training were C\# reimplementations of the DReCon and DeepMimic rewards and observations, adapted to a Unity3D environment. Further details on the training software architecture are provided in section \ref{arch} of the supplementary materials.

\subsection{Evaluation of the controller policies}

\subsubsection{Training strategy}
To find common ground for both pipelines, we chose to do all training with simulation steps every 0.008 s (125Hz), and up to 100M simulation steps. This took approximately 40 hours for each training run based on \emph{modular-agents}, and 40 minutes for each training run based on \emph{playground}.
All \emph{playground} training runs were performed on an Intel Core i9-14900K, 64\,GB of RAM, and two NVIDIA GeForce RTX 4090 graphics cards. All \emph{modular-agents} training runs were done either on a Windows~11 Pro with an Intel Core i9-14900K, 64\,GB of RAM, and a NVIDIA RTX 4070 Ti Super graphics card, or a Windows~11 Pro with an AMD Ryzen Threadripper 3970X and 128\,GB of RAM. The configuration files, including the environment and PPO hyperparameters, can be found in the supplementary materials. 

\subsubsection{How to compare different controller policies in VR }
Practical validation of the method included compilation for different platforms (Windows, Linux, MacOS, Android) to confirm the models could be used in production projects.
Since Android builds are not included in MuJoCo releases, we recompiled the physics engine from its source to integrate it into VR builds. We also compared these builds with running the entire environment in a web browser.

To compare the movement quality, independently of the controller policy used and the delivery method (Unity3D, WebXR), we recorded the movement data generated by the controller for exactly one animation loop, with no perturbations. We analysed the data with a set of custom Python scripts (which we provide in the supplementary materials). This allowed us to compare different training pipelines and controllers in the same context.

\subsubsection{How to compare different pipeline stages}

\label{sec:metrics}
To compare the motion generated by different policies with the reference animation used for training the policies, we used metrics with patterns that are easy to identify. For example, Figure~\ref{ee} shows the end-effector distance to the root for one reference animation, across five policies. In a walking pattern, the maxima and minima of this distance in the foot should be very well aligned with the reference animation.

To compare between policies, and between policies trained with different reference animations, we found that the measures common in PBCA literature did not seem to consider the perceptual studies on the topic (see section \ref{metrics}). We therefore complemented the standard metrics with an \emph{animation fidelity index}. 
Specifically, we considered three factors, with equal weight:

\begin{enumerate}
\item Instantaneous angular celerity, i.e., the magnitude of the angular velocity (one $\mathbb{R}$ value per joint)
\item Instantaneous linear celerity of end-effectors (hands, feet) relative to the root ($ 4 \times \mathbb{R}$ values per character)
\item Instantaneous linear celerity of the root, to reflect the displacement, independently of the forward direction ( one $ \mathbb{R}$ value per character)
\end{enumerate}

\begin{figure}[]
    \centering
    \includegraphics[width=0.5\textwidth]{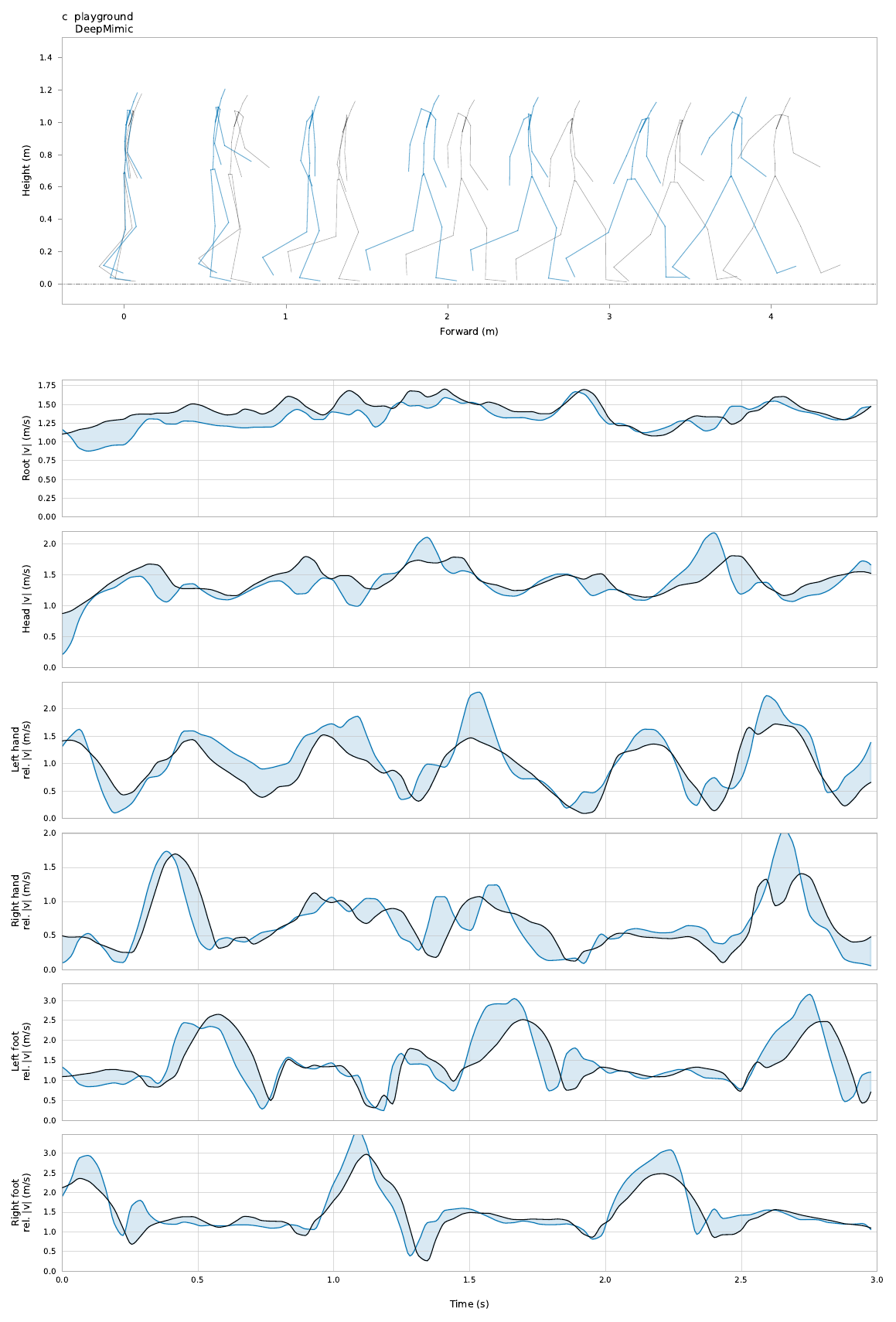}
\caption{A visualization of the metrics contributing to \emph{afi},
for the 
``Happy" motion of the \emph{LL} character.
Top: a sagittal stick-figure snapshot on metric scale, for illustration.  
Below, in order: the root linear-speed magnitude, contributing to the first term of the \emph{afi} metric;
the head celerity $|v_\mathrm{head}|$, one of the 19 celerities used for the second term of the \emph{afi} metric;
the four end-effector celerities, relative to the root, contributing to the third term of the \emph{afi} metric.
Shaded areas mark the per-frame deviation between reference and result and are the integrand of the cross-correlation used to calculate the  \emph{afi} scores reported in Table~\ref{table1:fid_scores}.}
\label{figure_one_anim}
\end{figure}

To derive a compact index from these metrics, we built a per-frame signal vector for each sequence and computed the cross-correlation between two sequences of poses---generated either by a reference animation or by a physics-based controller. 
Our animation fidelity index \emph{afi} was defined as the weighted sum of these: 

\begin{align}
 afi = \frac{1}{3}  * (\frac{\sum_{n=0}^{L} r_{ref}(n)r_{phy}(n+m)}{ \left\| r_{ref} \right\|^2 \left\| r_{phy} \right\|^2}\\
 \quad +   \frac{\sum_{n=0}^{L} a_{ref}(n)a_{phy}(n+m)}{ \left\| a_{ref} \right\|^2 \left\| a_{phy} \right\|^2}
\\
+  \frac{\sum_{n=0}^{L} e_{ref}(n)e_{phy}(n+m)}{ \left\| e_{ref} \right\|^2 \left\| e_{phy} \right\|^2} )
\end{align}



where:

\begin{compactitem}
   \item [r(n)] is the root linear celerity($ 1 \  \mathbb{R} $).
   \item [a(n)] is the angular celerity for all joints ($ 19 \ \mathbb{R} $).   
    \item [e(n)] is the celerity for end-effectors relative to the root ($ 4 \  \mathbb{R} $).
    \item [$_{ref}$] indicates the poses from the animation.
    \item [$_{phy}$] indicates the poses from the physical controller.
    \item [$L$] indicates the temporal length of the animation.
    \item [$m=0$] because we assume signals are temporally aligned.
\end{compactitem}
\ \\
The
\emph{afi}  is composed of three sub-indices. Term (1) reflects differences in speed of global movement. Term (2) reflects differences in the magnitude of angular velocities. 
Term in (3) reflects differences in the magnitude of end-effector velocities, relative to the root of the skeleton. 
To further illustrate the evaluation methodology, Figure~\ref{figure_one_anim} shows the metrics contributing to the \emph{afi} for the ``Happy'' walk of character ``LL'', trained in \emph{playground} with the \emph{DeepMimic} policy.

This index is inspired by perceptual studies  (see section \ref{metrics}): it is based on angular velocities, it takes into account the importance of the end-effectors and root displacement, it does \emph{not} compare pose rotations, and it is indifferent to the direction of the movement.

It is also universal, in the sense that it does not impose any constraint on physical contacts, or animation duration. Therefore, even if here we focus on walking motions,  it should also work for other tasks such as jumps, rolls, or grabbing objects. 
Moreover, since the per-joint angular celerity, the end-effector relative celerities, and the root linear celerity are non-negative by construction, the \emph{afi} stays within $[0,1]$, thus facilitating an intuitive interpretation as a similarity score.
We therefore use this metric as a convenient way to monitor the quality of the behavior generated by the controllers at different stages.

\section{Results}

\begin{table}[t]
    \centering
    \caption{Our three \emph{playground} controller policies, computed on the same 48 CPU rollouts (eight walks per cell). 
    \emph{MAE} is the mean absolute joint-angle error~\cite{Serifi:2024:VMP}; \emph{MPJPE} is the root-relative mean per-joint position error~\cite{Peng:2018:DeepMimic}; \emph{NPSS} is the normalised power-spectrum similarity~\cite{gopalakrishnan2019neural}; \emph{afi} is the animation fidelity index defined in section \ref{sec:metrics}. Bold marks the best score per column. The four metrics rank the methods identically (\emph{DeepMimic}~$>$~\emph{ADD}~$>$~\emph{AMP}) on both characters.}
    \label{table_MAE}
    \small
    \begin{tabular}{l|cc|cc|cc|cc}
    \toprule
            & \multicolumn{2}{c|}{MAE $\downarrow$ (\textdegree)} & \multicolumn{2}{c|}{MPJPE $\downarrow$ (mm)} & \multicolumn{2}{c|}{NPSS $\downarrow$} & \multicolumn{2}{c}{\emph{afi} $\uparrow$} \\
    Method & MM & LL & MM & LL & MM & LL & MM & LL \\
    \midrule
    DeepMimic   & \textbf{2.6} & \textbf{3.2} & \textbf{1.5} & \textbf{1.3} & \textbf{3.8} & \textbf{4.5} & \textbf{0.90} & \textbf{0.87} \\
    ADD         & 3.8 & 4.2 & 2.4 & 2.3 & 5.9 & 4.6 & 0.85 & 0.82 \\
      AMP         & 5.6 & 6.0 & 12.8 & 9.9 & 6.9 & 10.9 & 0.71 & 0.69 \\
    \bottomrule
    \end{tabular}
\end{table}

\subsection{Training results}
  All controllers evaluated achieve 100\% survival, both in Unity3D and WebXR, across all 16 motions. 
In Table \ref{table_MAE}, we show these metrics for our three \emph{playground} controller policies.
  \emph{ADD} produced movement with scores very close to \emph{DeepMimic} trained in \emph{playground}. We believe \emph{ADD} gives better  \emph{afi} scores than  \emph{AMP} because the displacement-based discriminator provides a more direct tracking signal.  
These findings were unexpected, as we anticipated that the results would align with the broader pattern observed in the literature:
 explicit tracking rewards~\cite{peng_deepmimic:_2018, Serifi:2024:VMP, Grandia:2024:BDX} consistently produce higher single-motion fidelity than adversarial approaches~\cite{peng_amp_2021}, a trade-off also highlighted by Li et al.~\cite{Li:2025:FeatureVsGAN}. 
Our results suggest that, while the two are not equivalent, the difference for \emph{ADD} is relatively small.

Adversarial methods (\emph{AMP, ADD}) offer practical advantages---no per-component reward tuning, implicit style capture, more controller flexibility---but at a measurable fidelity cost. \emph{ADD} narrows this gap substantially compared to \emph{AMP}, but \emph{DeepMimic} remains uniformly superior; for production scenarios requiring faithful reproduction of specific motions, the use of explicit rewards like in \emph{DeepMimic} remains the most reliable approach. However, when several animations need to be combined more flexibly, \emph{ADD} seems to provide a viable option.

In \emph{ADD} \cite{Zhang:2025:ADD},  the discriminator saturates rapidly (reaching near-100\% accuracy within 20M steps, since the zero-displacement positive sample is trivially separable), after which the style reward gradient vanishes. We emphasize that this saturation is a \emph{fundamental architectural limitation} of the displacement-based formulation, not a tuning problem: the discriminator's positive class is a single point (the zero vector), which is trivially separable from any non-zero displacement distribution regardless of network capacity, regularization, or training schedule. A small hybrid task reward component (5--15\% weight) 
resolves this by maintaining gradient flow after discriminator saturation.\textbf{}
With this fix, all 16 motions achieved 100\% survival in Unity3D and WebXR deployment.

\emph{afi} gives the same results as the joint-space (\emph{MAE}) and Cartesian (\emph{MPJPE}) metrics on the overall ranking: \emph{DeepMimic} is the strongest tracker on both skeletons, \emph{ADD} is the second-best, and \emph{AMP} trails on every column. The two metric families are complementary. \emph{MAE} and \emph{MPJPE} are pose-error metrics computed at the same frame index: they answer ``how far is the agent's pose from the reference's pose at time~$t$'' and are easy to inflate by a small phase shift, even when the trajectories are otherwise faithful. \emph{afi} is a time-aware cross-correlation of velocity and end-effector trajectories: it tolerates small phase shifts but penalises wrong overall pacing, missing follow-through, and jitter.
\emph{NPSS}, the frequency-domain reference metric, gives similar results, but it can sometimes be challenging 
because \emph{ADD} rollouts that terminate early concentrate spectral power at $f\to0$ -- the same low-frequency band some animation occupies.

\subsection{Losses in Animation Fidelity}

 \begin{figure}
  \centering
   \includegraphics[width=0.8\columnwidth]{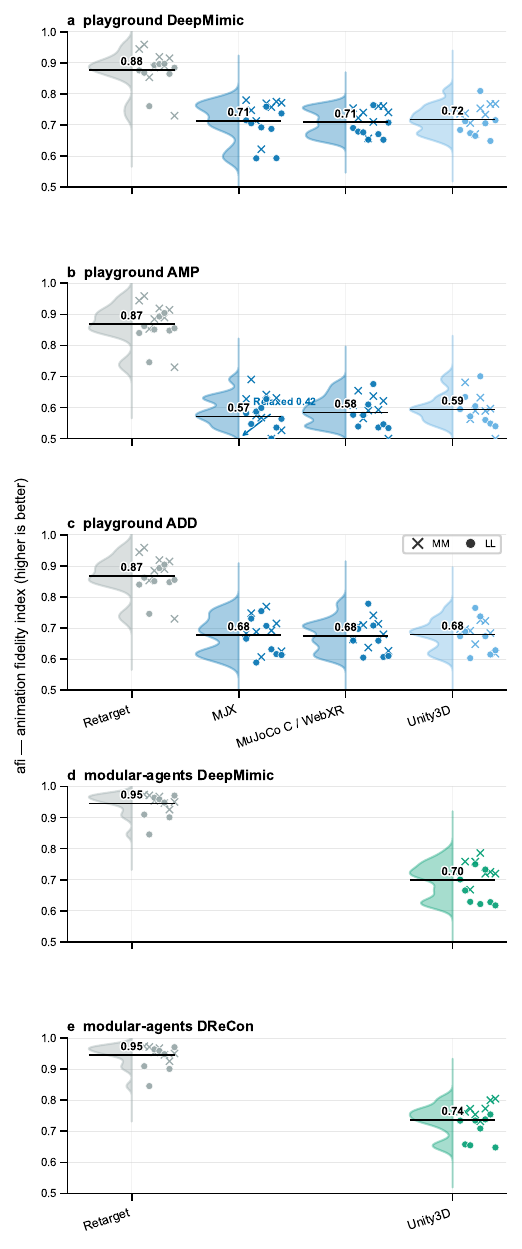}
\caption{Animation fidelity (\emph{afi}) at each pipeline stage. The \emph{Retarget} stage (FBX$\rightarrow$rig), in grey, shows the losses due to retargeting. 
The three \emph{playground} controllers, in blue, show the losses at different stages of the processing pipeline (rig$\rightarrow$MuJoCo~C (CPU)$\rightarrow$Unity3D).
The in-browser MuJoCo-WASM (\emph{WebXR}) rendering runs the same physics as the MuJoCo~C stage: its \emph{afi} is numerically identical to that column.
\ \\
At the bottom, the two \emph{modular-agents} baselines (green) only fill the \emph{Retarget} and Unity3D slots, since the training and inference environments are Unity3D. In this case, the \emph{Retarget} step conflates the retargeting always applied by Unity3D's Mecanim class, plus the retargeting introduced to make the reference MuJoCo ragdoll match the pose of the animated skeleton.
}
\label{fig4:fid_losses}
 \end{figure}

To better understand where animation fidelity is lost, Figure~\ref{fig4:fid_losses} shows the \emph{afi} at each stage of the \emph{playground} pipeline (MJX $\rightarrow$ MuJoCo~C $\rightarrow$ Unity3D), across animations. For comparison, we also show the losses in the \emph{modular-agents} pipeline.
\begin{enumerate}
  \item \textbf{Motion retargeting.} Virtually all methods for training physical controllers with reference motion data use a retargeting strategy to map kinematic poses, 
  defined as instantaneous rotations of a reference skeleton, to a targeted ragdoll. This step defines the targeted rotations and angular velocities in a reduced space defined by the humanoid's articulations. 
  To minimize the loss of accuracy of this step, we manually retargeted the captured animations to the reference characters and made the ragdolls match the proportions of the character skeletons
   as closely as possible. However, in both pipelines, there is minimal retargeting, from the rotations defining the kinematic poses to the reduced space of positions and velocities defined by the degrees of freedom of the articulated joints.
  This introduces a small but measurable loss.

\item \textbf{Sim2sim transfer.} To take advantage of GPU-accelerated training, our Playgorund pipeline involves transferring controllers trained in an MJX-based simulation to a MuJoCo~C-based simulation. 
These two MuJoCo versions handle contacts differently and operate on different hardware, potentially introducing noise and behavioral differences. We therefore tuned parameters across implementations manually to minimize these effects.
 As shown in Figure~\ref{fig4:fid_losses}, any remaining differences are now negligible
 (see Sections~\ref{training2} and the supplementary appendix~\ref{arch} for additional details on the mitigation process).

\item \textbf{Simulation environment.} Finally, the \emph{playground} pipeline involves transferring the controllers from a Python-based to a Unity3D-based environment.
The underlying engine is in both cases MuJoCo~C, so theoretically there shouldn't be a difference, but the codebase is still different and any discrepancy could lead to different results. 
Figure \ref{fig4:fid_losses} shows that this difference is negligible, but it was significant during the process of establishing the processing pipeline and debugging the controller policies. We believe it is important to monitor this and any other possible source of difference.

\end{enumerate}

\subsection{Comparing policies}

Table~\ref{table1:fid_scores} summarizes the values of the animation fidelity index for each controller and animation reference rendered in Unity3D. 
 To compare different training policies, we performed Wilcoxon paired tests, taking $p<0.01$ as a threshold for a significant difference.
All training policies have scores significantly different ($p<0.0001$) from the reference (which has by construction a score of 1 for any animation). 

The comparison of \emph{DReCon} and \emph{DeepMimic} policies implemented in \emph{modular-agents} confirms that, everything else being equal, having explicit information of the targeted pose at every frame produces better movement than only receiving rewards on movement similarity (mean difference $0.038, p<0.001$). This comes, of course, at the cost of having to load and process the animation data at the inference stage.

The comparison of \emph{DeepMimic} policies trained in \emph{modular-agents} and \emph{playground} suggests that, everything else being equal, the \emph{playground} pipeline gives equivalent results in terms of animation fidelity (mean difference $0.018, p=0.27$). This suggests that the faster GPU pipeline enabling 100M+ environment steps in minutes does not sacrifice quality. 

In this regard, where \emph{DReCon} has the best score of the \emph{modular-agents} controllers, \emph{DeepMimic} outscores all other \emph{playground} controllers (\emph{AMP} shows a mean difference of $0.143, p <0.0001$; \emph{ADD} shows a mean difference of $0.058, p = 0.0021$).
If we focus only on \emph{playground} controllers, we observe that \emph{ADD} is a clear improvement over \emph{AMP} (mean difference of $0.085, p<0.0001$), but still does not match \emph{DeepMimic} (mean difference $0.038, p=0.0042$). Other metrics lead to similar conclusions. In supplementary section \ref{anim_fidelity_metrics}, we show that the more common mean absolute joint-angle error (MAE) leads to similar conclusions when comparing playground policies.

 \begin{table}

\caption{Animation fidelity
for each controller vs.\ its reference at Unity3D deployment (\emph{afi} $ \in [0,1]$, \emph{higher is better}). Because \emph{afi} is a normalised cross-correlation, it is independent of the animation length (Sec.~\ref{sec:metrics}); it is computed over the full recorded loop, not a fixed window.
\textbf{Bold} is the best (highest) score per row. Embedded violins show the distribution behind each column ($\times$=MM, $\bullet$=LL, bar=mean).}

\label{table1:fid_scores}
\centering
 \includegraphics[width=\linewidth]{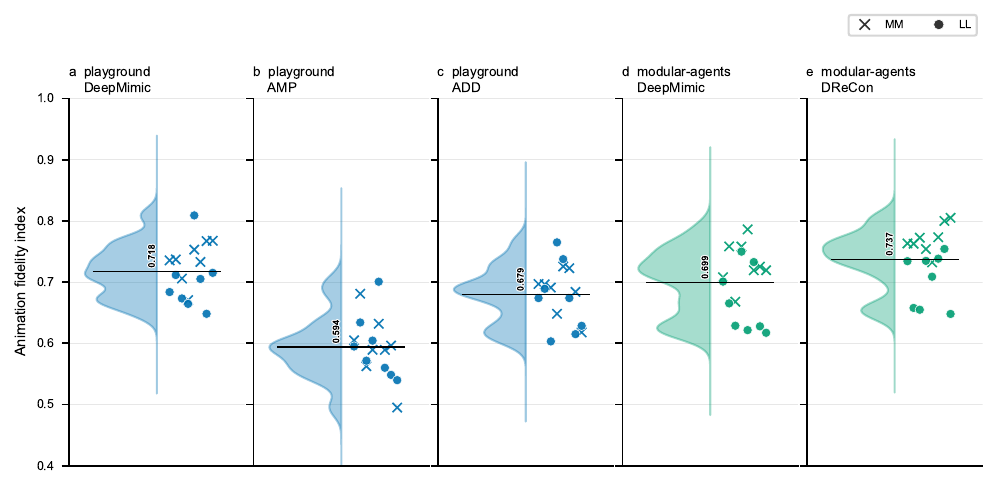}
\small
{\setlength{\tabcolsep}{5pt}%
\begin{tabular}{ll cc ccc}
\toprule
&  & \multicolumn{3}{c}{\textbf{playground}} & \multicolumn{2}{c}{\textbf{modular-agents }}\\
\cmidrule(lr){3-5} \cmidrule(lr){6-7}
& \textbf{Style}   & DeepMimic & AMP & ADD  & DeepMimic & DReCon \\
\midrule
 LL & Neutral    & 0.684 & 0.595 & 0.674 & 0.701 & \textbf{0.734}  \\
 & Angry     & \textbf{0.712} & 0.634 & 0.689   & 0.665 & 0.658 \\
 & Drag       & \textbf{0.673} & 0.572 & 0.603 & 0.629 & 0.655   \\
 & Fashion    & 0.664 & 0.604 & \textbf{0.765} & 0.750 & 0.735  \\
 & Happy      & \textbf{0.809} & 0.701 & 0.738 & 0.622 & 0.709  \\
 & Relaxed   & 0.705 & 0.560 & 0.674  & 0.733 & \textbf{0.738}  \\
 & Sad        & 0.648 & 0.548 & 0.615 & 0.628 & \textbf{0.754}  \\
 & Wounded    & \textbf{0.715} & 0.540 & 0.629 & 0.617 & 0.648  \\
\midrule
 MM & Neutral    & 0.736 & 0.605 & 0.697 & 0.708 & \textbf{0.763}  \\
 & Angry      & 0.737 & 0.681 & 0.697 & 0.758 & \textbf{0.763}  \\
 & Drag       & 0.706 & 0.563 & 0.691 & 0.668 & \textbf{0.773}  \\
 & Fashion    & 0.671 & 0.589 & 0.648 & \textbf{0.758} & 0.754  \\
 & Happy     & 0.753 & 0.632 & 0.726  & \textbf{0.786} & 0.732  \\
 & Relaxed   & 0.733 & 0.589 & 0.723  & 0.719 & \textbf{0.773}  \\
 & Sad       & 0.767 & 0.596 & 0.684  & 0.725 & \textbf{0.800}  \\
 & Wounded    & 0.768 & 0.495 & 0.618 & 0.719 & \textbf{0.805}  \\
\midrule
 \multicolumn{2}{l}{\textbf{Average (all)}}  & 0.718 & 0.594 & 0.679 & 0.699 & \textbf{0.737}  \\
\bottomrule
\end{tabular}}
 \end{table}

\subsection{Visual inspection}
We complemented the quantitative metrics with expert feedback for the best \emph{playground} policies (i.e. DeepMimic and ADD). The professional animator who retargeted and cleaned the animations (co-author \textbf{[removed for submission anonymity]}
) did a systematic review comparing the animations in Motion Builder, with the controller policies in the web-based environment.  In supplementary section \ref{expert_eval}, a detailed breakdown can be found for each animation
used to train the \emph{playground} policies \emph{DeepMimic} and \emph{ADD}. 

Even the best policies moved too much the chest, which also made the arms and the head move more than in the reference animation. Sometimes the policy introduced jolts towards the sides, as well as exaggerated movements, which then had implications for the arms and the head. 
We believe that this is due to the policy training for stability -i.e., not falling-, but the phenomenon is found across policies, and we have not found a systematic way to fix it. 

This result may have appeared previously  in the PBCA literature, but it may be less apparent due to the use of simpler skeletons (particularly with less joints on the spine), and to not using rigged and textured characters.

\section{Discussion}

\subsection{Losses in retargeting}

The automated FBX-to-MJCF pipeline successfully produced training-ready ragdolls for both the adult male (MM) and female child (LL) skeletons, despite their significantly different proportions. The key step was the retargeting: the IK-based retargeting we used preserved end-effector positions well, but may have introduced artifacts in the torso chain---particularly for the LL skeleton. 
To map skeleton poses to the reduced space of the physical ragdoll, we still did automated retargeting, as have done virtually all recent contributions to PBCA.

An alternative strategy, specific to our use case, could be to calculate the  $qPos$ and $qVel$ needed to train the physical controllers simply using finite differences and joint constraints.
This would only be reliable and precise because of the match between the skeleton poses and the ragdoll topology and it could, possibly, further reduce the fidelity loss introduced by the retargeting, even if it our results suggest it was already quite small. 

\subsection{Sim to sim transfer}

Transferring controllers from MuJoCo-MJX to MuJoCo~C  introduced a measurable but small fidelity gap. 
This is analogous to the sim-to-real transfer challenges documented in robotics~\cite{He:2025:ASAP, Serifi:2025:Thesis}, and our observation-design approach complements the domain randomization strategies typically used in that context~\cite{Serifi:2024:VMP}.
We had expected this to be the main source of degradation in the MJX-based pipeline, but our results suggest that, in the final implementation of our \emph{playground} pipeline, the degradation was minimal. The critical mitigation strategy was excluding simulator-specific observations (composite inertia tensors) from the policy input. The small remaining gap was probably dominated by differences in how collisions are handled, which affect contact-rich phases of the gait cycle.

\subsection{Controller losses}

The main animation fidelity loss was introduced in the training step. 
It could be argued that this was due to the constraint of having physically possible movements. 
However, we took all possible steps to preserve the animation of physical movements 
(creating characters matching the proportions of the actors, minimal editing of the animation). 
We also chose to work with walking motions because they do not involve too many physical contacts, and can easily be created in different styles.  
The different controller policies varied in the movement quality they produce, but the results of \emph{DeepMimic} and \emph{ADD} do suggest that there is a 
general threshold that physical controller policies do not manage to breach. 





\subsection{Physics-based controllers for production} 
\label{pbca4prod}

We found that showing the result on rigged and textured characters is important to judge the quality of the motion generated by controller policies,  but
we could not find any article in the PBCA literature where the poses produced by the controller policy are directly mapped on a rigged and textured character. The recent MotionBricks \cite{wang2026motionbricksscalablerealtimemotions} integrated controller policies in Unreal Engine, but the mapping to a rigged and textured character was done with retargeting and inverse kinematics, which did not allow validating the motion quality produced by the controller policies independently of the rest of the processing steps.
It should also be pointed that  MotionBricks also joins a growing amount of work \cite{Luo:2023:PHC,Luo:2024:PULSE, Luo:2025:SONIC, Juravsky:2024:SuperPADL, Liao:2025:BeyondMimic} where the datasets and infrastructure needed for training the controller policies go largely beyond what is available in a typical VR production. This makes it difficult to adopt these controller policies for VR production.

After inspecting the controller policies applied on rigged and textured characters, we believe there remains a fundamental gap in animation fidelity that is not solved with current policies. 
This animation fidelity gap is hard to detect with the most commonly used success metrics, but we could readily identify it comparing side by side the reference animation used for training and the movement generated by the policies trained, particularly when using animations with minimal physical constraints, like the walking cycles we used. For visual comparison,  the deployment of the controllers in the VR-compatible web environment was very useful, and we've made it available together with the training pipeline.\footnote{The source code of the web environment used is included in the supplementary materials and can also be found in   
\textbf{[link removed for submission anonymity]}.} 

The ceiling of the \emph{afi} scores on the \emph{playground} policies, which stayed far lower from the \emph{afi} scores after retargeting, might also reflect this animation fidelity gap. 
In figure \ref{fig4:fid_losses}, it can be seen in the main \emph{afi} score loss for \emph{playground} policies, which occurs at the training stage, even before transferring to MuJoCo~C.
Further evaluation with improved policies should validate whether the ceiling found in \emph{afi} scores really reflects the limits in animation fidelity found through visual inspection. 

Overall, we believe this gap is a fundamental challenge caused by the way these policies are trained: at the early stages of training the trials where the character does not fall produce a longer training episode, and therefore higher rewards. The strong movements of the chest help considerably to keep balance, and are therefore learnt early, even if these movements are not found in the reference animation. As the training progresses, the imitation rewards of DeepMimic and the adversarial comparison of \emph{AMP} and \emph{ADD} progressively bring the movement closer to the reference, but they do not overwrite these early learnt movements. The elimination of simulator-specific observations, such as inertia tensors, might have been good for transferring policies from MJX to MuJoCo~C, but it might also contribute to this gap.

\section{Conclusions and future work}

Adversarial controllers like \emph{ADD} provide results close to methods requiring explicit pose tracking (\emph{DReCon}) or explicitly tracking the phase of an animation (\emph{DeepMimic}), 
which encourages exploration of their use in production projects involving richer interactive behavior. However, they struggle with very asymmetric movements (like the \emph{Wounded} animation, where the \emph{ADD} policy fails to move forward in both characters). A possible solution is to introduce complementary rewards to compensate for such problems, and doing so specifically for challenging animations.

When motion quality is an important goal, we recommend matching ragdoll and skeleton proportions and showing the result on rigged and textured characters, with a one-to-one retargeting of the ragdoll controller to the rigged and textured character, rather than ragdoll visualizations alone. 
Visual inspection is significantly more sensitive to movement errors on rigged and textured characters---subtle artifacts in joint velocities, foot skating, and phase shifts that are invisible on a ragdoll become immediately apparent on a rigged and textured mesh (see also supplementary video). This practice would help raise the bar for animation quality in physics-based control research.
In addition, rendering PBCA controllers in WebXR can give animation fidelity results comparable to rendering them in a Unity3D environment. Moreover, doing so is very useful for a visual comparison of the movement generated by the policy and the reference animation. 
We recommend adopting systematic visual comparison with these tools. The metrics commonly used in the physics-based literature do not capture movement anomalies that an animation expert can readily detect. We recommend complementing any objective metric with expert assessment.

Systematically assessing animation fidelity at different stages of transferring the controller to the final environment is necessary to detect quality losses. This includes not only the well explored sim2sim or sim2real transfer challenge, but also any retargeting and change of the environment where the simulation is run.
 This may help to improve movement quality, but also to understand why challenging tasks, involving rich contacts or complex interactions, are not learnt effectively. 
 For the reference motion, if any motion retargeting is needed, manual methods are highly recommended.

 Despite all these considerations, we could not reliably generate physics-based controller policies for VR characters matching the quality of the reference animation. The behavior generated by the trained policies does allow identifying the reference motion, but even the best training methods induce the controller to move much more the chest than in the reference animation. There remains a significant gap between the reference animation and the movement generated by the controller policy. 
 We believe this gap is difficult to close because of the way the controller policies are trained. 

Future work should explore alternative training strategies resulting in PBCA policies that  bridge this animation fidelity gap and generate movement closer to the reference animation data. 
We would also like to complement this work by extending the evaluation to motions involving richer contacts and object interaction~\cite{Xu:2025:InterMimic},
 and to integrate multi-motion universal controllers~\cite{Serifi:2024:VMP, Luo:2024:PULSE, tessler2024maskedmimic} into the comparison.

\section*{Supplementary Materials}
\begin{compactitem}
  \item [S1] Supplementary text further detailing the architecture decisions, more details on objective evaluation metrics and details on the expert evaluation
  \item [S2] Supplementary video
  \item [S3] link to code repository \textbf{[removed for submission anonymity]}
  \item [S4] link to data repository \textbf{[removed for submission anonymity]} 
\end{compactitem}

\section*{Author Contributions}

\textbf{[removed for submission anonymity]}

\section*{Acknowledgments}

\textbf{[removed for submission anonymity]}

\bibliographystyle{abbrv-doi}

\bibliography{references}

\clearpage
\appendix


This document accompanies the main paper ``Animation fidelity in physics-based
humanoid controllers''. It collects architecture and training decisions,
additional animation-fidelity discussion, and complementary analyses.

\section{Architecture and training decisions}

\label{arch}


\begin{figure}[]
  \centering
      \includegraphics[width=8cm]
      {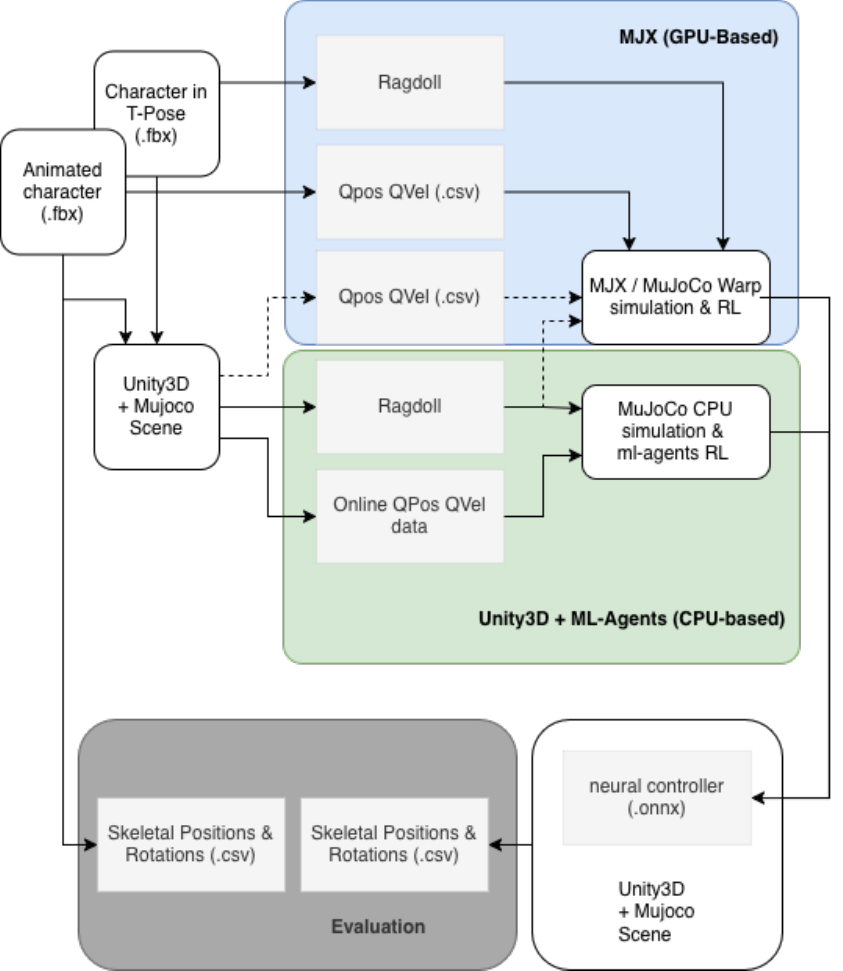}
    \caption{We evaluate the animation fidelity produced by two different training pipelines, compared to the reference animation.}
    \label{fig1:two}
  \end{figure}

A significant effort of this work has been finding the right combination of software packages that enable seemless transfer from an MJX training pipeline to a Unity3D-based inference.
Figure \ref{fig1:two} shows a high-level architecture of the two pipelines and how we collected per-frame poses of the controller to compare with the reference motion.

However, this does not show specific software package choices.  We chose ONNX runtime because the default package to load ONNX models in Unity3D (Sentis) has practical differences in implementation that make the resulting control actions quite different. Therefore, only a port of ONNX Runtime enabled appropriate inference.\footnote{Sentis and ONNX Runtime may diverge on the order of $10^{-2}$ per output dimension, which the neural network controller will probably amplify.}

More generally, we selected MuJoCo \emph{playground} over IsaacGym or IsaacLab for two reasons:
(i) MuJoCo \emph{playground} shares its physics engine with the MuJoCo~C runtime
used in our Unity3D deployment efforts (via the Unity3D \texttt{mujoco-unity} plugin),
so the sim-to-sim gap reduces to numerical (CPU vs.\ XLA on GPU) rather than
algorithmic differences in contact resolution;
(ii) MuJoCo \emph{playground} uses MJX, which is fully JAX-traceable,
which lets us co-locate the environment, the PPO update,
and the discriminator (for \emph{AMP/ADD}) under a single \texttt{jit} boundary.
The trade-off is that asset import in MJX is currently more restrictive
than IsaacLab's USD pipeline, which is why we author skeletons in MJCF
and convert to Unity3D at deployment time rather than the reverse.

For \emph{modular agents}, we used a stable PD with a controller action every two steps. For MJX training, we used traditional PD controllers. This simplified transfer. However, we found that having slightly less frequent controller updates reduced jitter. We therefore chose to use control updates every 4 steps.
\ \\
One practical consequence of the 4-substep control window is that
the reference trajectory must be re-sampled to the 31\,Hz control rate before training, which truncates the original 120\,Hz motion capture acquisition frequency to the
nearest multiple of four substeps.
Depending on the duration of the animation loop, this can imply drops $\leq 4$ frames at the seam---small in absolute terms,
but enough to introduce a sub-frame discontinuity that \emph{AMP}'s discriminator
amplifies. Future work should resample
the reference inside the environment with a continuous-time interpolant
rather than at asset-loading time, which would remove the seam artefact and, possibly, 
could improve \emph{AMP}.

\begin{table}

\caption{Animation fidelity as mean per-joint orientation error (degrees, \emph{lower is better}), for each controller vs.\ its reference at Unity3D deployment. \textbf{Bold} is the best (lowest) score per row. $^{\dagger}$\emph{modular-agents} DeepMimic diverges (heading drift) on LL Wounded; the $127^{\circ}$ failure is kept in the average. Embedded violins show the distribution behind each column ($\times$=MM, $\bullet$=LL, bar=mean).}

\label{table3:fid_scores}
\centering
 \includegraphics[width=\linewidth]{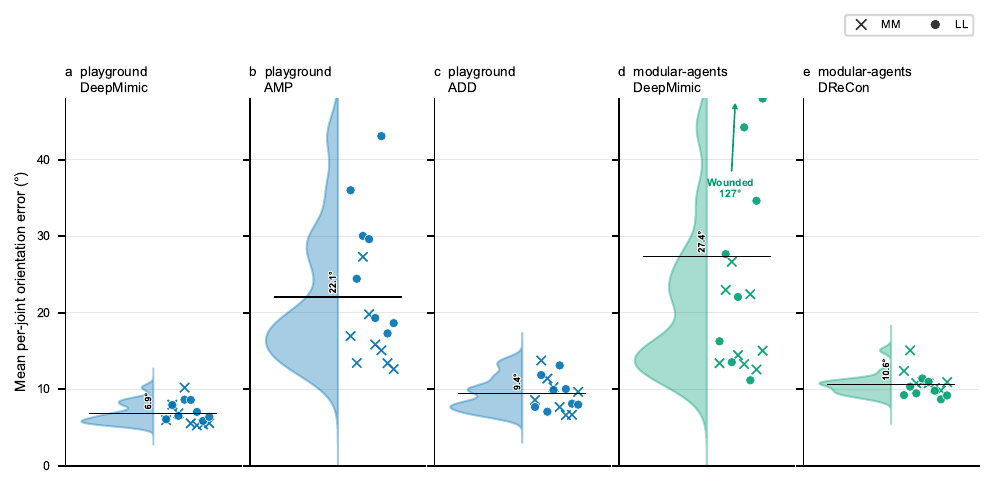}
\small
{\setlength{\tabcolsep}{5pt}%
\begin{tabular}{ll cc ccc}
\toprule
&  & \multicolumn{3}{c}{\textbf{playground}} & \multicolumn{2}{c}{\textbf{modular-agents }}\\
\cmidrule(lr){3-5} \cmidrule(lr){6-7}
& \textbf{Style}   & DeepMimic & AMP & ADD & DeepMimic & DReCon  \\
\midrule
 LL & Neutral     & \textbf{6.1} & 36.0 & 7.7 & 16.3 & 9.2   \\
 & Angry           & \textbf{7.9} & 24.4 & 11.9 & 27.7 & 10.3  \\
 & Drag           & \textbf{6.5} & 30.0 & 7.1 & 13.5 & 9.5   \\
 & Fashion        & \textbf{8.6} & 29.6 & 9.9 & 22.1  & 11.4  \\
 & Happy           & \textbf{8.6} & 19.3 & 13.1 & 44.2 & 11.0  \\
 & Relaxed         & \textbf{7.0} & 43.1 & 10.0 & 11.2 & 9.8  \\
 & Sad             & \textbf{5.9} & 17.3 & 8.1 & 34.6 & 8.7  \\
 & Wounded         & \textbf{6.4} & 18.6 & 8.0 & 127.3$^{\dagger}$ & 9.2  \\
\midrule
 MM & Neutral     & \textbf{6.0} & 16.9 & 8.6  & 13.4 & 12.4  \\
 & Angry          & \textbf{8.0} & 13.4 & 13.8 & 23.0 & 15.1   \\
 & Drag          & \textbf{6.9} & 27.3 & 11.4  & 26.7 & 10.8   \\
 & Fashion         & \textbf{10.2} & 19.8 & 10.3 & 14.5 & 11.0  \\
 & Happy           & \textbf{5.5} & 15.8 & 7.7 & 13.3 & 10.8  \\
 & Relaxed         & \textbf{5.3} & 15.1 & 6.6 & 22.4 & 10.2  \\
 & Sad             & \textbf{5.4} & 13.4 & 6.7 & 12.6 & 9.8  \\
 & Wounded         & \textbf{5.6} & 12.6 & 9.7 & 15.0 & 10.9  \\
\midrule
 \multicolumn{2}{l}{\textbf{Average (all)}}  & \textbf{6.9} & 22.1 & 9.4 & 27.4 & 10.6  \\
\bottomrule
\end{tabular}}
 \end{table}

 \emph{AMP} is particularly sensitive to the quality of the reference motion data: loop-seam discontinuities in the reference cause the discriminator to learn from physically impossible velocity spikes, and near-static motions (e.g., a subtle limp) cause temporal aliasing where the discriminator cannot form a meaningful decision boundary.
Both failure modes can be addressed through hybrid reward (combining a small explicit tracking component with the adversarial objective), confirming the findings of Li et al.~\cite{Li:2025:FeatureVsGAN} that explicit rewards complement adversarial methods.

\section{Animation fidelity metrics}
\label{anim_fidelity_metrics}

Table~\ref{table3:fid_scores} reports the same 16 controller/reference pairs scored with the mean per-joint orientation error, the pose-error metric most commonly used in the tracking literature. Table \ref{table3:fid_scores} suggests that in \emph{playground} \emph{DeepMimic} has the best score, followed closely by \emph{ADD}. When comparing \emph{modular-agents} trainings, it also shows DReCon is better than DeepMimic, as expected.
 The two metrics agree on the \emph{playground} ranking (\emph{DeepMimic}~$>$~\emph{ADD}~$>$~\emph{AMP}) but disagree on the \emph{modular-agents} baselines: because it is a per-frame pose error, it is inflated by the small phase shifts that \emph{afi} tolerates, and it is dominated by a single diverging rollout (\emph{modular-agents} \emph{DeepMimic} on LL Wounded, $127^{\circ}$).

\section{Expert evaluation}
\label{expert_eval}
Table \ref{table_expert} details the problems found when looking at the movement generated by the policies for each of the reference animations used to train the policies. The visual inspection was limited 
 to the \emph{playground} policies with better scores (\emph{DeepMimic} and \emph{ADD}) 

\begin{table*}

\caption{Expert evaluation of the \emph{playground} policies \emph{DeepMimic} and \emph{ADD}, for each animation used to train them (translated from french). The subjective score given was between 0 - bad and 100 - good}

\label{table_expert}
\centering
\small
{\setlength{\tabcolsep}{2pt}%
\begin{tabularx}{18cm}{lllcC}
\toprule
&  & & Score & Problems detected  \\
\midrule
 MM & DeepMimic & Neutral     & 
 80	& The chest moves, which has repercussions on the joints further down on the hierarchy; arms and head. It gives the impression that the character has a limp, maybe due to a non-constant step size. Slight footskating visible.  \\
 & & Angry           & 75	& 
 The chest moves, which has repercussions on the joints further down on the hierarchy; arms and head. It gives the impression that the character has a limp.  \\
 & & Happy           & 75	&The chest moves a lot, with a jolt towards the left, which affects the joints further down the hierarchy; arms and head. Slight impression of limping, little footskating. \\
 & & Relaxed        & 85	& The chest moves a lot more than the reference, which affects the joints further down the hierarchy; arms and head. However, in a slow movement this is less disturbing. Slight footskating visible. \\
 & & Wounded           & 80	& The possible instabilities are not disturbing since the movement is already quite unstable. The speed of the step when it moves the wounded foot is slower than the reference.  \\
 & & Drag         & 80 &	The chest moves, which has repercussions on the joints further down on the hierarchy; arms and head. It gives the impression that the character has a limp, maybe due to a non-constant step size. Slight footskating visible.  \\
 & & Sad             & 85 &	The chest moves more than it should, compared to the reference, which has repercussions on the joints further down on the hierarchy; arms and head. But in a slow movement this is less disturbing. Slight footskating visible. \\
 & & Fashion         & 75 &	The chest moves, which has repercussions on the joints further down on the hierarchy; arms and head. It gives the impression that the character has a limp, maybe due to a non-constant step size. Slight footskating visible. \\
\midrule
 MM & ADD & Neutral     & 
 70	& The chest moves a lot, which affects the joints further down the hierarchy; arms and head. The pose is too bent backwards, compared to the reference. Slight footskating visible.  \\
 & & Angry           & 60	&
 The chest moves a lot, and in a way that is not stable during the cycle, something that affects the joints further down in the hierarchy; arms and head. The pose is too bent backwards, compared to the reference. Slight footskating visible.  \\
 & & Happy           & 75	&The chest moves a lot, with a jolt towards the left, which affects the joints further down the hierarchy; arms and head. Slight impression of limping, little footskating. \\
 & & Relaxed        & 75	& The chest moves more than it should, compared to the reference, which has repercussions on the joints further down on the hierarchy; arms and head. But in a slow movement this is less disturbing. Footskating well visible \\
 & & Wounded           & 10	& He does not manage to move forward   \\
 & & Drag         & 70 &	The chest moves a lot, and in a way that is not stable during the cycle, something that affects the joints further down in the hierarchy; arms and head. The pose is too bent backwards, compared to the reference. Footskating visible.  \\
 & & Sad             & 75	& The chest moves a lot more than the reference, which affects the joints further down the hierarchy; arms and head. However, in a slow movement this is less disturbing.  Footskating clearly visible. \\
 & & Fashion         & 65	& The chest moves a lot, which affects the joints further down the hierarchy; arms and head. The pose is too bent backwards, compared to the reference. Footskating visible.e \\

\midrule
 LL & DeepMimic & Neutral     & 
 80	& The chest moves, which has repercussions on the joints further down on the hierarchy; arms and head. It gives the impression that the character has a limp, maybe due to a non-constant step size. Slight footskating visible \\
 & & Angry           & 75	& The chest moves a lot, which has repercussions on the joints further down on the hierarchy; arms and head. Slight footskating visible  \\
 & & Happy           & 70 &	The chest moves, which has repercussions on the joints further down on the hierarchy; arms and head. Slight footskating visible. The joint in the middle of the foot does not seem to be animated. \\
 & & Relaxed        & 60	& The chest moves  a lot with strange jolts on the right, which affects the joints further down the hierarchy, legs and head. Footskating very visible. The reference already gives a strange impression, the spine is twisted, so it is not a good starting point.\\
 & & Wounded           & 80	& The possible instabilities are not disturbing since the movement is already quite unstable. Left foot wounded is not constant because it drags on the floor. Slight footskating  \\
 & & Drag         & 75	& The chest moves, which has repercussions on the joints further down on the hierarchy; arms and head. Slight footskating visible.   \\
 & & Sad             & 80& The chest moves, which has repercussions on the joints further down on the hierarchy; arms and head. Slight footskating visible.  \\
 & & Fashion         & 60	& The chest moves  a lot with strange jolts, which affects the joints further down the hierarchy; legs and head. Footskating  visible and the position of the foot does not follow the reference (where the feet are one in front of the other on a line). \\

\midrule
 LL & ADD & Neutral     & 
 75	& The chest moves  a lot, which affects the joints further down the hierarchy, legs and head.  Slight footskating visible. \\
 & & Angry           & 65 &	The chest moves  a lot, a jolt towards the left, which affects the joints further down the hierarchy; legs and head. Footskating  visible. \\
 & & Happy           & 50	& The chest moves  a lot, position unnatural with jolts, which affects the joints further down the hierarchy; legs and head. Footskating visible.\\
 & & Relaxed        & 45	& The chest moves a lot with strange jolts towards the right, which affects the joints further down the hierarchy; legs and head. Footskating very visible. The reference gives  a strange impression, the spine is twisted so it is not a very good starting point.\\
 & & Wounded           & 10	& She does not manage to move forward  \\
 & & Drag         & 70	& The chest moves a lot, and in a way that is not stable across the animation cycle, which affects the joints further down the hierarchy, arms and head. Footskating visible.   \\
 & & Sad             & 55	& The chest moves a lot, and in a way that is not stable through the animation cycle, which affects the joints further down the hierarchy, arms and head. Footskating visible. \\
 & & Fashion         & 50	& The chest moves  a lot with strange jolts, which affects the joints further down the hierarchy; legs and head. Footskating  visible and the position of the foot does not follow the reference (where the feet are one in front of the other on a line). \\

\bottomrule
\end{tabularx}}
 \end{table*}

\end{document}